\documentclass[conference]{IEEEtran}
\IEEEoverridecommandlockouts
\usepackage{multirow,graphicx}
\usepackage{subcaption}
\usepackage{amsmath,amssymb,amsfonts}
\usepackage{algorithm}
\usepackage{algpseudocode}
\usepackage{booktabs}
\usepackage{tabularx}
\usepackage{array}
\usepackage{textcomp}
\usepackage{textcomp,verbatim,xcolor,soul,tikz}
\usepackage{cite}
\usepackage{hyperref}
\usepackage{soul}

\def\BibTeX{{\rm B\kern-.05em{\sc i\kern-.025em b}\kern-.08em
    T\kern-.1667em\lower.7ex\hbox{E}\kern-.125emX}}

\begin{document}

\title{ESN-DAGMM: A Lightweight Framework for Unsupervised Time-Series Data Monitoring in 5G O-RAN Networks\\
}

\author{
\IEEEauthorblockN{Andrew J Chen}
\IEEEauthorblockA{
\textit{andrewcsd66@gmail.com}}
\and
\IEEEauthorblockN{Raymond Zhao}
\IEEEauthorblockA{
\textit{zraymond@vt.edu}}
\and
\IEEEauthorblockN{Lingjia Liu}
\IEEEauthorblockA{
\textit{ljliu@vt.edu}}

\thanks{A. Chen is a high school student enrolled in Canyon Crest Academy.}
\thanks{R. Zhao and L. Liu are with Wireless@Virginia Tech, Bradley Department of Electrical and Computer Engineering, Virginia Tech. The work of R. Zhao and L. Liu are partially supported by the National Telecommunications and Information Administration (NTIA)'s Public Wireless Supply Chain Innovation Fund under Award No. 51-60-IF012. Any opinions, findings, and conclusions or recommendations expressed in this publication are those of the authors and do not necessarily reflect the views of the NTIA.}
\thanks{Accepted for publication in the proceedings of the IEEE International Conference on Data Mining (ICDM) 2025 UGHS Symposium.}
\thanks{© 2025 IEEE.  Personal use of this material is permitted.  Permission from IEEE must be obtained for all other uses, in any current or future media, including reprinting/republishing this material for advertising or promotional purposes, creating new collective works, for resale or redistribution to servers or lists, or reuse of any copyrighted component of this work in other works.}
}

\maketitle

\begin{abstract}
Open Radio Access Network (O-RAN) is an important 5G network architecture enabling flexible communication with adaptive strategies for different verticals. 
However, testing for O-RAN deployments involve massive volumes of time-series data (e.g., key performance indicators), creating critical challenges for scalable, unsupervised monitoring without labels or high computational overhead. To address this, we present ESN-DAGMM, a lightweight adaptation of the Deep Autoencoding Gaussian Mixture Model (DAGMM) framework for time series analysis. Our model utilizes an Echo State Network (ESN) to efficiently model temporal dependencies, proving  effective in O-RAN networks where training samples are highly limited. Combined with DAGMM's integratation of dimensionality reduction and density estimation, we present a scalable framework for unsupervised monitoring of high volume network telemetry. When trained on only 10\% of an O-RAN video-streaming dataset, ESN-DAGMM achieved on average 269.59\% higher quality clustering than baselines under identical conditions, all while maintaining competitive reconstruction error. By extending DAGMM to capture temporal dynamics, ESN-DAGMM offers a practical solution for time-series analysis using very limited training samples, outperforming baselines and enabling operator's control over the clustering-reconstruction trade-off.
\end{abstract}


\section{Introduction}
Open Radio Access Network (O-RAN) represents a transformative shift in wireless infrastructure design, offering flexibility, cost savings, and improved performance for 5G, 6G and beyond through disaggregated, interoperable components \cite{ShadabO-RAN25}. While these deployments provide significant advantages, ensuring reliable performance requires continuous monitoring of network conditions. O-RAN's management and automation challenges mean that effective control and monitoring are essential for maintaining network performance~\cite{ORAN_Challenge}. 
In this setting, unsupervised structure discovery methods, such as clustering or density modeling, can help operators understand network states without task-specific labels.

However, scalability remains a critical challenge, since testing specifications cover a wide range of services from video streaming to web browsing. Each service requires monitoring of numerous key performance indicators (KPIs), creating a rapidly growing volume of data \cite{oran-e2e}. This data deluge increases communication overhead between RAN components, creating a bottleneck that could impede widespread O-RAN adoption if not addressed. Practical methods should therefore (i) reduce dimensionality to a latent representation while preserving temporal information and (ii) pair that representation with an interpretable density model that can reveal structure to support monitoring. To address these requirements, we propose a lightweight model designed for KPI time series data. While not evaluated here for anomaly detection (AD), the framework provides a flexible foundation that can support such tasks. 

Traditional methods for AD and structure discovery, such as statistical tests~\cite{KianAD20}, clustering algorithms \cite{puclusteringdetection}, and Principal Component Analysis (PCA) \cite{10.1145/1269899.1254895}, are valued for their simplicity and minimal computational cost. These approaches, including rule-based statistical detection and clustering, effectively isolate outliers in low-dimensional settings by reducing dimensionality via techniques like PCA \cite{clustering}. 

Yet these methods face limitations with modern, complex datasets. For example, statistical approaches assume distributions rarely met in practice, while clustering struggles when anomalies lack separability. Additionally, PCA’s linear model and reliance on feature engineering hinder it in high-dimensional, nonlinear data \cite{panganomalyreview, 10.1145/1269899.1254895}. 

To address these issues, deep learning has emerged as a strong alternative. Such methods learn representations directly from high-dimensional data, detecting subtle anomalies that traditional models often miss. Common approaches include reconstruction models, such as autoencoders, which learn normality and use reconstruction error to flag deviations. Another approach uses prediction models like Recurrent Neural Networks (RNN), which forecast values and identify anomalies based on prediction error  \cite{deeplearning,deeplearning2,deeplearning3,panganomalyreview}.

Scaling anomaly detection models become challenging when training and inference costs increase sharply with data volume. To mitigate this, many methods first map high-dimensional data into a compact latent representation. This approach, which often combines reconstruction models with clustering or probabilistic estimators, has led to state-of-the-art performance by balancing expressiveness and interpretability. For example, methods have been used to learn video embeddings for clustering \cite{QIU2024110550} or to cluster pose graphs \cite{DBLP:journals/corr/abs-1912-11850}.

We build upon the Deep Autoencoding Gaussian Mixture Model (DAGMM), a framework that handles high-dimensional data by mapping it into a latent space for effective AD \cite{zong2018deep}. In DAGMM, an autoencoder generates latent representations and reconstruction error. An estimation network then uses these features to output the soft membership of each data point to a finite number of Gaussian components.
End-to-end training enables direct anomaly scoring in the latent space. Although effective for static datasets, DAGMM's original design overlooks temporal dependencies. This limitation is a significant drawback for analyzing time-series data, which necessitates an encoder capable of efficiently capturing temporal structure.



To address this, we introduce ESN-DAGMM, a lightweight adaptation of DAGMM for time-series data. Our approach replaces the original MLP-based autoencoder with a recurrent one that leverages an Echo State Network (ESN) as the encoder. ESNs are a scalable choice because their lightweight design fixes reservoir weights and trains only the output layer \cite{jaeger2001echo}. Their proven effectiveness in wireless communications \cite{jxmimoofdm,sjxai,sjconfigured} makes them ideal for our model. ESN-DAGMM simultaneously maps high-dimensional KPI time-series data to a latent space and fits a GMM for interpretable clustering, preserving temporal structure. This method reduces data complexity, enhances temporal modeling, and supports lightweight inference, with key contributions including a time-series adapted DAGMM and demonstrated superior clustering performance for scalable O-RAN monitoring under highly constrained training scenarios.

In summary, our contributions are:
\begin{itemize}
    \item We introduce a time-series adaptation of the DAGMM framework that combines an ESN encoder with an estimation network, enforcing a Gaussian structure in the latent space.
    \item We show through experiments that our framework achieves higher clustering performance than baseline approaches while maintaining competitive reconstruction error, making it a practical option for scalable KPI monitoring in O-RAN.
\end{itemize}

The rest of this paper is organized as follows. We first present the problem formulation in Section~\ref{sec:problem_formulation}, then describe our methodology in Section~\ref{sec:methodology}. We follow with experimental results in Section~\ref{sec:results} and conclude in Section~\ref{sec:conclusion}.
\section{Problem Formulation and Methodology}

\subsection{Problem Formulation}
\label{sec:problem_formulation}

Consider a time-series dataset \(\mathbf{X} = \{\mathbf{x}_1, \mathbf{x}_2, \dots, \mathbf{x}_T\}\), where each observation \(\mathbf{x}_t \in \mathbb{R}^D\) represents a set of \(D\) metrics captured at time \(t \in \{1, \dots, T\}\). This data often exhibits complex, non-linear dynamics. Our objective is to map \(\mathbf{X}\) to a compact representation \(\mathbf{z_c} \in \mathbb{R}^d\), where \(d \ll T \times D\). At the same time, we aim to enforce a probabilistic structure on this space, specifically modeling it as a mixture of \(K\) Gaussian functions, to enhance interpretability and facilitate clustering tasks or AD.

\begin{figure}[htbp]
    \centering
    \includegraphics[width=1.0\linewidth]{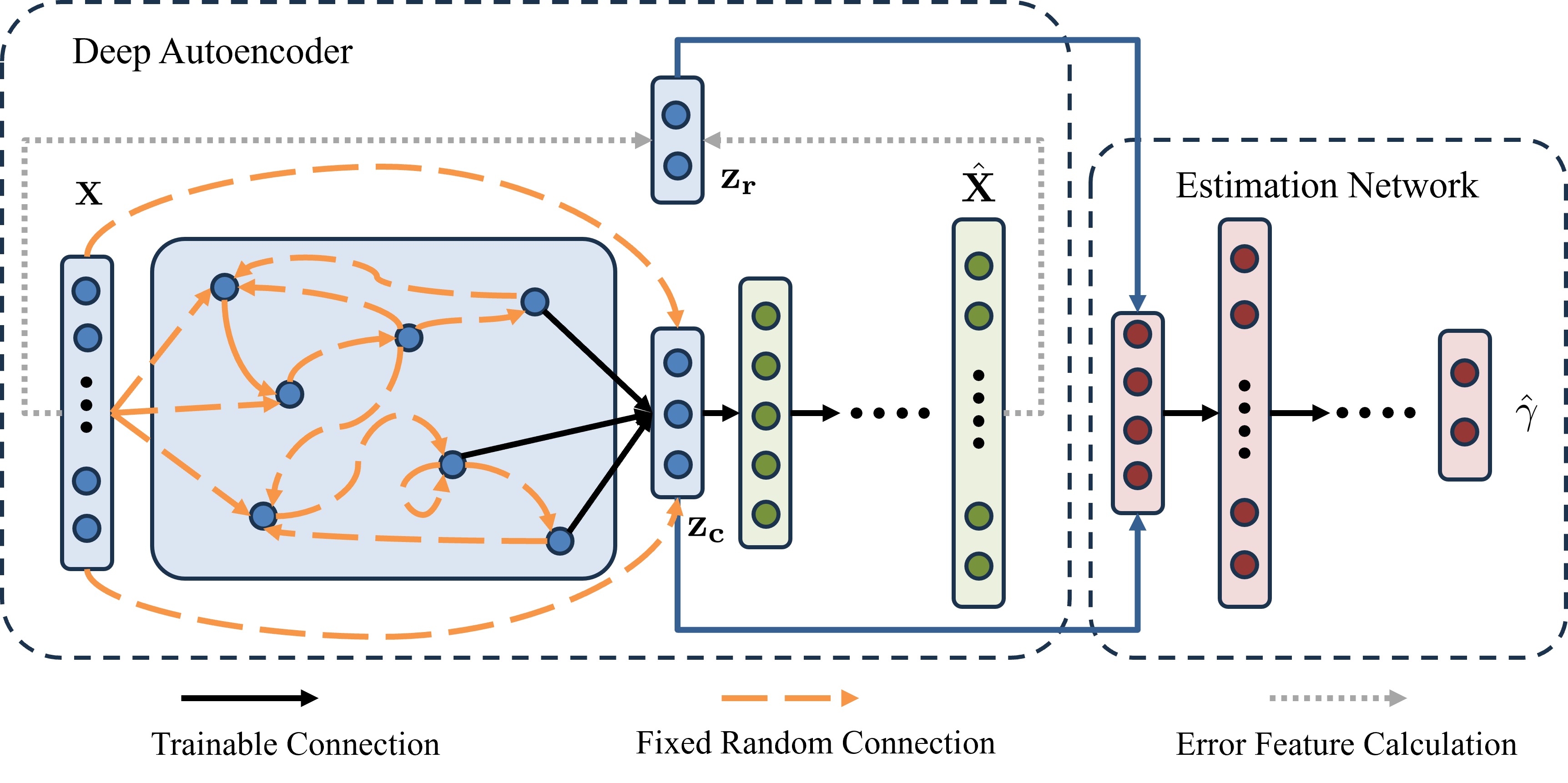}
    \caption{DAGMM architecture for time-series data.}
    \label{fig:dagmm_arch}
\end{figure}

To achieve this dual objective, we adapt the DAGMM framework \cite{zong2018deep} for sequential data processing. Figure~\ref{fig:dagmm_arch} illustrates its architecture, consisting of an encoding, decoding, and estimation network. The encoder and decoder, parameterized by $\theta_e$ and $\theta_d$ respectively, perform dimensionality reduction and reconstruction. Specifically, the encoder $h(\cdot)$, maps the input time-series $\mathbf{X}$ to a compact latent code \(\mathbf{z_c} \in \mathbb{R}^{d}\). The decoder $g(\cdot)$, then reconstructs the input \(\hat{\mathbf{X}} \approx \mathbf{X}\) from $\mathbf{z}_c$. These operations are formally defined as:
\begin{align}
\mathbf{z}_c &= h(\mathbf{X}; \theta_e), \\
\hat{\mathbf{X}} &= g(\mathbf{z}_c; \theta_d),
\end{align}

To augment the latent representation with information about reconstruction quality, a feature vector \(\mathbf{z}_r \in \mathbb{R}^m\) is derived from the input and its reconstruction: 
\begin{equation}
\mathbf{z}_r = f(\mathbf{X}, \hat{\mathbf{X}}),
\end{equation}
where \(f(\cdot)\) computes aggregated reconstruction error features. The final latent vector fed into the estimation network is then \(\mathbf{z} = [\mathbf{z_c}; \mathbf{z}_r]\), such that \(\mathbf{z} \in \mathbb{R}^{d+m}\).

The estimation network, parameterized by $\theta_m$, processes $\mathbf{z}$ to predict soft mixture memberships, $\hat{\gamma}$, representing the probability of a sample belonging to each of $K$ Gaussian components. For a batch of \(N\) sequences, the Gaussian mixture parameters are estimated as:
\begin{equation}
    \hat{\phi}_k = \frac{1}{N} \sum_{i=1}^N \hat{\gamma}_{ik},
\end{equation}
\begin{equation}
    \hat{\boldsymbol{\mu}}_k = \frac{\sum_{i=1}^N \hat{\gamma}_{ik} \mathbf{z}_{i}}{\sum_{i=1}^N \hat{\gamma}_{ik}},
\end{equation}
\begin{equation}
\hat{\boldsymbol{\Sigma}}_k = \frac{\sum_{i=1}^N \hat{\gamma}_{ik} (\mathbf{z}_{i} - \hat{\boldsymbol{\mu}}_k)(\mathbf{z}_{i} - \hat{\boldsymbol{\mu}}_k)^T}{\sum_{i=1}^N \hat{\gamma}_{ik}},
\end{equation}
for each component \(k = 1, \dots, K\), where \(\hat{\phi}_k\), \(\hat{\boldsymbol{\mu}}_k\), and \(\hat{\boldsymbol{\Sigma}}_k\) are the mixture weight, mean, and covariance, respectively.

The training of the DAGMM involves optimizing a joint objective function \(\mathcal{J}(\theta_e, \theta_d, \theta_m)\), which integrates three loss components over \(N\) sequences to balance dimensionality reduction, structure, and stability:
\begin{equation}
\label{eq:joint_objective}
    \mathcal{J} = \frac{1}{N} \sum_{i=1}^N L(\mathbf{X}_i, \hat{\mathbf{X}}_i) + \frac{\lambda_1}{N} \sum_{i=1}^N E(\mathbf{z}_{i}) + \lambda_2 P(\hat{\boldsymbol{\Sigma}}),
\end{equation}
where hyperparameters \(\lambda_1\) and \(\lambda_2\) control the trade-off.

The objective function consists of three components:
\begin{itemize}
    \item Reconstruction loss (\(L(\cdot,\cdot)\)) quantifies the autoencoder's performance by minimizing the mean squared error between the input and its reconstructed counterpart, averaged over the sequence, thus preserving temporal patterns.
    \begin{equation}
      L(\mathbf{X}_i, \hat{\mathbf{X}}_i) = \frac{1}{T} \sum_{t=1}^T \|\mathbf{x}_{i,t} - \hat{\mathbf{x}}_{i,t}\|_2^2,
    \end{equation} 
    
    \item GMM sample energy loss (\(E(\mathbf{z})\)) models the probabilities of observing input samples within the learned mixture model, and aims to find the optimal network combination that maximizes the likelihood of observing these samples.
    \begin{equation}
        \begin{split}
          E(\mathbf{z}) = \\ -\log \Bigg( \sum_{k=1}^{K} \hat{\phi}_k \frac{\exp\left(-\frac{1}{2}(\mathbf{z} - \hat{\boldsymbol{\mu}}_k)^T \hat{\boldsymbol{\Sigma}}_k^{-1} (\mathbf{z} - \hat{\boldsymbol{\mu}}_k)\right)}{\sqrt{(2\pi)^{d+m} |\hat{\boldsymbol{\Sigma}}_k|}} \Bigg)
        \end{split}
  \end{equation}

  \item Covariance matrix regularization (\(P(\hat{\boldsymbol{\Sigma}})\)) prevents the GMM's singularity problem, where trivial solutions can arise from degenerate covariance matrix diagonal entries, by penalizing small values.
  \begin{equation}
      P(\hat{\boldsymbol{\Sigma}}) = \sum_{k=1}^K \sum_{j=1}^{d+m} \frac{1}{\hat{\Sigma}_{kjj}},
  \end{equation}
  
\end{itemize}

This end-to-end optimization yields a structured latent vector that not only effectively performs dimensionality reduction on the high-dimensional input but also inherently embeds a probabilistic prior. This combination enhances both the compactness and interpretability of the learned representations.

\subsection{Methodology}
\label{sec:methodology}

Our pipeline leverages specific neural network architectures to implement the DAGMM framework, focusing on dimensionality reduction and analysis of high-dimensional time-series data. It integrates an ESN as the encoder, an MLP as the decoder, and a similarly MLP-based estimation network to impose a GMM structure on the latent space.

The ESN-based encoder is used due to its efficacy in capturing temporal dependencies within time-series data while offering remarkable computational efficiency. An ESN operates by maintaining a high-dimensional, sparsely connected recurrent "reservoir" state that evolves over time. The key to its efficiency lies in its fixed, randomly initialized reservoir weights, meaning only the output layer weights are trained. This results in fewer trainable parameters and computational operations compared to fully trainable recurrent networks.

The state update equation for the reservoir at each time step \(t\) is given by:
\begin{equation}
\mathbf{s}_t = (1 - \alpha) \mathbf{s}_{t-1} + \alpha \cdot \tanh(\mathbf{W}_{in} \mathbf{x}_t + \mathbf{W}_{res} \mathbf{s}_{t-1}),
\end{equation}
where \(\mathbf{s}_t \in \mathbb{R}^{d_{res}}\) is the reservoir state, \(\alpha\) is the leaky rate, \(\mathbf{W}_{in} \in \mathbb{R}^{d_{res} \times D}\) is the input weight matrix, \(\mathbf{W}_{res} \in \mathbb{R}^{d_{res} \times d_{res}}\) is the recurrent weight matrix, and \(\tanh(\cdot)\) is the nonlinear activation. After processing the full sequence, the latent vector is derived from a readout layer:
\begin{equation}
\mathbf{z_c} = \mathbf{W}_{out} [\mathbf{s}_T; \text{flatten}(\mathbf{X})],
\end{equation}
where \(\mathbf{W}_{out} \in \mathbb{R}^{d \times (d_{res} + T \cdot D)}\) is the trainable output weight matrix, and \(\text{flatten}(\mathbf{X})\) concatenates the input sequence. This design allows the ESN to efficiently capture and condense the complex dynamics of high-dimensional time-series data.

The decoder is implemented as 4-layer MLP. Its role is to reconstruct the original time-series from the compact latent vector learned by the ESN encoder. This MLP decoder directly maps the latent representation to the full reconstructed space, with each layer utilizing a $\text{ReLU}$ activation function to introduce non-linearity. 

To provide comprehensive information for the GMM, the reconstruction quality is incorporated into the latent dimension. Our feature vector $\mathbf{z}_r$ aggregates reconstruction errors, namely mean square error (MSE), relative Euclidean distance, and cosine similarity between the input and reconstruction. 

The estimation network is implemented as a 3-layer MLP to process this latent vector to predict soft mixture memberships. This network incorporates dropout layers between its fully connected layers and the final output layer utilizes a softmax activation function to provide a probability distribution over the $K$ Gaussian components. Given vector $\mathbf{z}$, this network outputs a prediction:
\begin{equation}
    \mathbf{p}=MLP(\mathbf{z}; \theta_m), \qquad \hat{\gamma} = \text{softmax}(\mathbf{p}),
\end{equation}
where $\mathbf{p}$ is the raw output from the MLP, and $\hat{\gamma}$ is the soft assignment described in Section \ref{sec:problem_formulation}. 

Based on these assignments, the GMM parameters ($\hat{\phi}_k, \hat{\boldsymbol{\mu}}_k,\hat{\boldsymbol{\Sigma}}_k$) are estimated. The entire model is then optimized using the joint objective function \eqref{eq:joint_objective}, balancing reconstruction accuracy, probabilistic structuring of the latent space, and model stability through covariance regularization. This architecture effectively combines the ESN's temporal processing with the MLP's reconstruction capabilities and the GMM's interpretability for robust time-series analysis.
\section{Results Presentation}
\label{sec:results}

The dataset utilized in this study was collected from a $2 \times 2$ MIMO O-RAN testbed that used software-defined radios to emulate a realistic communication environment with random OFDM burst interference. This setup, consisting of a base station (BS), user equipment (UE), a video streaming server, and a C++-based interferer, logged measurements every 20 ms over 120-second runs. Data was gathered over several days to create a robust, high-dimensional time-series dataset.

The dataset contains a comprehensive set of physical layer and medium access control (PHY/MAC) KPIs, packet captures, and video statistics, providing a detailed representation of system behavior under interference. As a preprocessing step, we standardized the 13 network KPIs using a standard scaler to ensure fair comparison across features. The dataset, accessible at \cite{orandataset}, was also prepared by applying a windowing technique where each sample consists of 28 past time steps, with each step containing all 13 network KPIs.



To evaluate the performance of our DAGMM framework, we employed two primary metrics:
\begin{itemize}
    \item \textbf{Reconstruction MSE}: Measures the average reconstruction error across the 13 standardized features, with lower values indicating better reconstruction accuracy.
    \item \textbf{Silhouette Score (SS)}: Evaluates clustering quality in the latent space, ranging from $-1$ to $+1$, where higher values indicate well-separated and cohesive clusters.
\end{itemize}

For a comprehensive performance comparison, we evaluated our approach against several baselines, each representing different strategies for dimensionality reduction and clustering:
\begin{itemize}
    \item \textbf{PCA + GMM}: A classical two-stage baseline where PCA performs linear dimensionality reduction, followed by fitting a GMM to the latent space.
    \item \textbf{ESN Autoencoder + GMM}: A two-stage approach leveraging an ESN-based autoencoder for temporal modeling, followed by GMM clustering.
    \item \textbf{MLP-DAGMM}: The original single-stage DAGMM framework implemented using standard MLPs.
    \item \textbf{LSTM-DAGMM}: A single-stage pipeline integrating a Long Short-Term Memory (LSTM) network as the encoder for recurrent processing.
    \item \textbf{RNN-DAGMM}: A single-stage pipeline using a standard RNN as the encoder for recurrent processing.
    \item \textbf{ESN-DAGMM}: Our single-stage pipeline, integrating an ESN as the encoder within the end-to-end DAGMM framework.
\end{itemize}

To simulate a practical scenario where training data is limited, as is common in highly dynamic wireless environments, we used only 10\% of the dataset for training, reserving the remaining 90\% for testing. This constrained setup was designed to evaluate model performance under challenging conditions, reflecting the difficulty of acquiring large, consistent datasets in rapidly changing network environments.



\begin{figure}[htbp]
    \centering
    \includegraphics[width=1.0\linewidth]{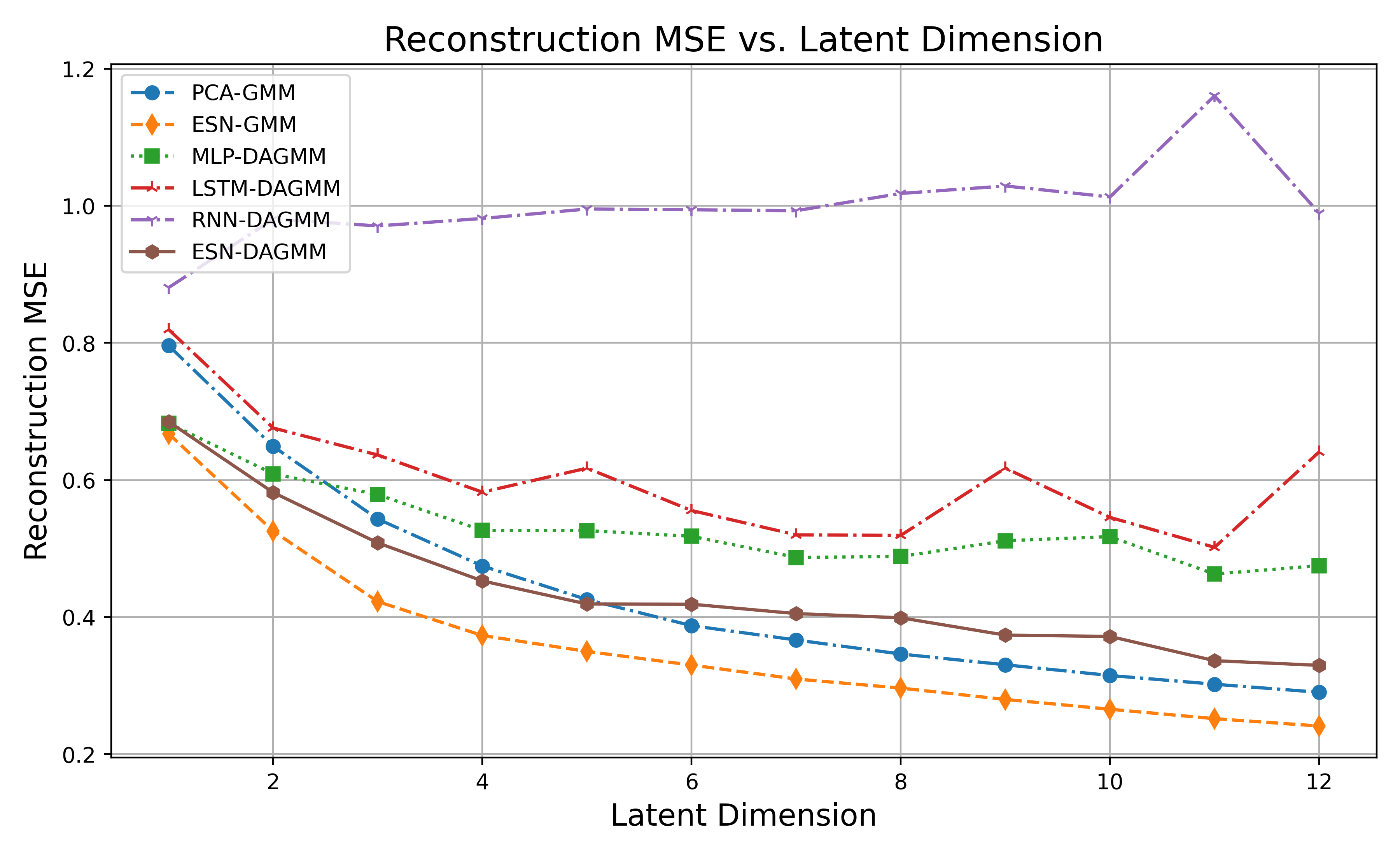}
    \caption{Reconstruction MSE as a function of latent dimension for various models.}
    \label{fig:dim_vs_mse}
\end{figure}

Reconstruction MSE as a function of the latent dimension is illustrated in Figure \ref{fig:dim_vs_mse}. As the latent dimension increases, nearly all models show a decreasing trend in MSE, indicating a larger space can capture more information from the original data. The two-stage baselines achieve the lowest MSE because they are optimized solely for dimensionality reduction and reconstruction. In contrast, the single-stage DAGMM models balance reconstruction and clustering; consequently, they generally have higher MSEs.

The models' performance in terms of MSE begins to saturate around higher dimensions. The gain in reconstruction accuracy, therefore, diminishes beyond a certain point. We selected a latent dimension of 10 for our analysis because the MSE for most models is around its lowest point. This dimensionality represents a significant reduction, as it maps each $28 \times 13$ data sample down to a $10 \times 1$ latent vector, while still retaining sufficient information to capture essential details from the original data. 

In this limited training scenario, our ESN-DAGMM approach performs competitively with the two-stage baselines and demonstrates a significantly lower MSE than other DAGMM frameworks like MLP, LSTM, and RNN. The ESN's architecture, which only trains the output layer weights, makes it more sample efficient than traditional recurrent networks that require training of all weights. This allows ESN-DAGMM to effectively learn a compact latent representation with only 10\% of data, outperforming the more complex and data-hungry recurrent models in this constrained scenario. The specific MSE values for each model at a latent dimension of 10 are detailed in Table \ref{tab:resultsd10k2}.

\begin{table}[htbp]
\centering
\caption{Performance comparison of ESN-DAGMM and baselines at 10 dimensions and 2 clustering components.}
\label{tab:resultsd10k2}
\resizebox{\columnwidth}{!}{%
\begin{tabular}{lcccc}
Model & \multicolumn{1}{l}{MSE} & \multicolumn{1}{l}{MSE \% Diff} & \multicolumn{1}{l}{Silhouette Score} & \multicolumn{1}{l}{SS \% Diff} \\ \hline
PCA-GMM & 0.315 & -18.1\% & 0.179 & 136.3\% \\
ESN-GMM & 0.266 & -39.8\% & 0.129 & 227.9\% \\
MLP-DAGMM & 0.517 & 28.0\% & 0.268 & 57.8\% \\
LSTM-DAGMM & 0.912 & 59.2\% & 0.411 & 2.9\% \\
RNN-DAGMM & 1.013 & 63.3\% & 0.131 & 222.9\% \\
ESN-DAGMM & 0.372 & --- & 0.423 & ---
\end{tabular}%
}
\end{table}


While two-stage models excelled in reconstruction, their clustering performance was inferior. Our ESN-DAGMM demonstrated superior clustering, as evidenced by its silhouette score. As detailed in Table \ref{tab:resultsd10k2}, ESN-DAGMM significantly outperformed all baselines when identifying two clusters at a latent dimension of 10. We focused on two clusters for this comparison because empirically, the baselines methods struggled to learn or identify more than two distinct groups. These results highlight the efficacy of our approach in learning distinct, well-separated clusters.

\begin{figure}[htbp]
    \centering
    \includegraphics[width=1.0\linewidth]{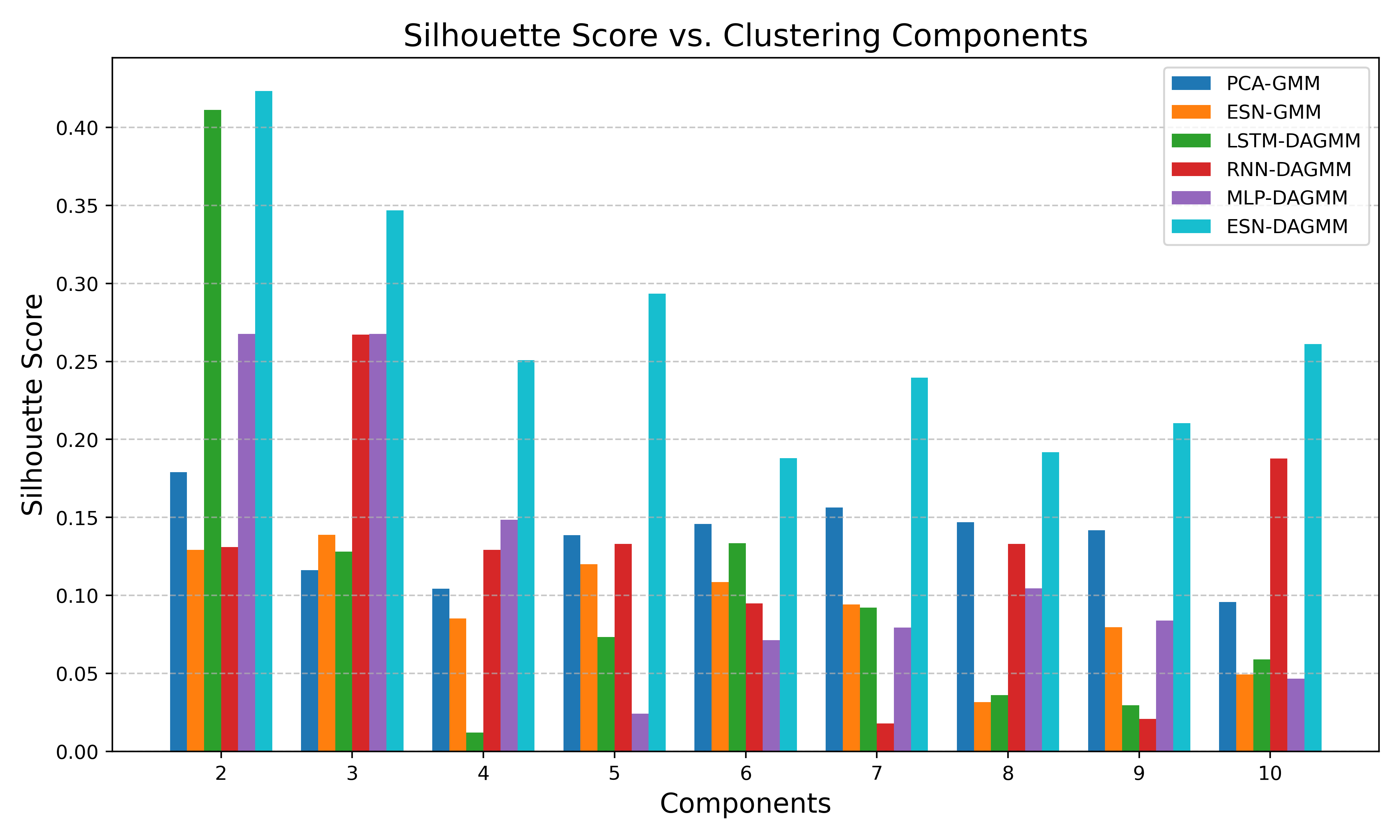}
    \caption{Silhouette scores showing clustering performance for different frameworks.}
    \label{fig:cluster_vs_ss}
\end{figure}

Clustering performance across varying numbers of Gaussian components (2 to 10) is shown in Figure \ref{fig:cluster_vs_ss}, where ESN-DAGMM consistently outperforms other models. On average, ESN-DAGMM outperforms PCA-GMM by \textbf{102.4\%}, ESN-GMM by \textbf{227.1\%}, LSTM-DAGMM by \textbf{449.4\%}, RNN-DAGMM by \textbf{310.4\%}, and MLP-DAGMM by \textbf{258.7\%}. Overall, ESN-DAGMM outperforms baselines by \textbf{269.59\%}.


A key advantage of ESN-DAGMM is its flexibility, allowing tuning of the \((\lambda_1, \lambda_2)\) parameters to prioritize either reconstruction or clustering, unlike two-stage approaches that optimize for one objective at the expense of the other. This adaptability enables a balanced optimization tailored to specific application needs. Empirically, \(\lambda_1\) was varied from 0.1 to 1.0 and \(\lambda_2\) from 0.001 to 0.01, with these ranges adjusted across different latent dimensions to optimize performance, reflecting the model's sensitivity to the trade-off observed.
\section{Conclusion}
\label{sec:conclusion}

This study presents an adaptation of the DAGMM framework for time-series analysis, incorporating an ESN as the encoder to process high-dimensional time-series data. Our ESN-DAGMM approach demonstrates robust performance in both dimensionality reduction and interpretable clustering. By leveraging the temporal modeling capabilities of ESN, the framework effectively captures essential features of the input data while achieving superior clustering performance, as evidenced by significant improvements in silhouette scores compared to baseline methods. The framework's adaptable design further enhances its versatility, allowing optimization for either reconstruction accuracy or clustering quality based on application requirements. These results highlight the potential of ESN-DAGMM as a versatile tool for analyzing complex time-series data in communication systems and beyond.

\bibliographystyle{IEEEtran}
\bibliography{IEEEabrv, ref}

\end{document}